\documentclass{PoS}

\title{The novel photon detectors based on MPGD technologies for the upgrade of
COMPASS RICH-1}

\ShortTitle{COMPASS Hybrid PDs}

\author{J.Agarwala$^a$, M.Alexeev$^b$, C.D.R.Azevedo$^c$, R.Birsa$^d$,
F.Bradamante$^e$, A.Bressan$^e$, C.Chatterjee$^e$, M.Chiosso$^b$,
A.Cicuttin$^a$, P.Ciliberti$^e$, M.L.Crespo$^a$, S.Dalla Torre$^d$, S.S.Dasgupta$^d$,
O.Denisov$^f$, M.Finger$^g$, M.Finger Jr.$^g$, B.Gobbo$^d$, M.Gregori$^d$,
G.Hamar$^d$, S.Levorato$^d$, A.Maggiora$^f$, 
A.Martin$^e$, G.Menon$^d$, J.Novy$^g$, D.Panzieri$^h$, F.A.B.Pereira$^c$,
C.A.Santos$^d$, G.Sbrizzai$^e$, M.Slunecka$^g$, K.Steiger$^i$,
L.Steiger$^i$, M.Sulc$^i$, \speaker{F.Tessarotto}$^d$\thanks{corresponding author},
J.F.C.A.Veloso$^c$, Y.Zhao$^d$\\
\llap{$^a$}INFN, Sezione di Trieste and Abdus Salam ICTP, Trieste, Italy\\
\llap{$^b$}INFN, Sezione di Torino and University of Torino, Torino, Italy\\
\llap{$^c$}I3N - Physics Department, University of Aveiro, Aveiro, Portugal\\
\llap{$^d$}INFN, Sezione di Trieste, Trieste, Italy\\
\llap{$^e$}INFN, Sezione di Trieste and University of Trieste, Trieste, Italy\\
\llap{$^f$}INFN, Sezione di Torino, Torino, Italy\\
\llap{$^g$}Charles University, Prague, Czech Republic and JINR, Dubna, Russia\\
\llap{$^h$}INFN, Sezione di Torino and University of East Piemonte, Alessandria, Italy\\
\llap{$^i$}Technical University of Liberec, Liberec, Czech Republic\\
        E-mail: \email{fulvio.tessarotto@ts.infn.it}}

\abstract{
The RICH-1 Detector of the COMPASS experiment at CERN SPS has undergone an important upgrade in 2016.
Four new photon detectors, based on MPGD technology and covering a total active area larger than 1.2~$m^2$
have replaced the previously used MWPC-based photon detectors.
The new detector architecture, resulting from a dedicated, eight years long, R\&D program, consists in a
hybrid MPGD combination of two THGEMs and a Micromegas stage; the first THGEM, coated with a CsI
layer, acts as a reflective photocathode. 
The signals are extracted from the anode pads by capacitive coupling and read-out by analog
front-end electronics based on the APV25 chip. 
The new COMPASS RICH-1 photon detectors are described in detail: the detector design,
the engineering aspects, the mass production, and the quality assessment are discussed.
The assembly of the MPGD components and the installation of the new detectors are illustrated together with
the main aspects of the commissioning. Preliminary indication of performance results are also presented.
}

\FullConference{5th International Conference on Micro-Pattern Gas Detectors (MPGD2017)\\
		22-26 May, 2017\\
		Philadelphia, USA}

\begin{document}

\section{Introduction}

The COMPASS RICH-1~\cite{Albrecht} detector is a large gaseous Ring Imaging Cherenkov Counter
providing hadron identification in the range of momenta between 3 and 60 GeV/c, over a large angular
acceptance ($\pm$200 mrad), for the COMPASS Experiment~\cite{COMPASS} at CERN SPS.

It consists of a 3~m long $C_{4}F_{10}$ radiator, a 21~$m^2$ large focusing VUV mirror surface
and Photon Detectors (PDs) covering a total active area of 5.5~$m^2$.
Three photodetection technologies are used in RICH-1:  Multi Wire Proportional Chambers (MWPCs) with CsI photocathodes,
Multi Anode Photo-Multipliers Tubes (MAPMTs) and Micro Pattern Gaseous Detectors (MPGDs) based PDs.

COMPASS RICH-1 was designed and built between 1996 and 2001  and is in operation since 2002.
The whole photodetection surface was originally equipped with MWPCs hosting 16 CsI-coated photocathodes
of about 600$\times$600 mm$^2$ active area.

Inspite of their good performance, MWPCs with CsI-coated photocathodes have limitations in terms of maximum
effective gain, time response, rate capability and aging of the CsI photoconverter.
In 2006, to cope with the high particle rates of the central region, 4 of the 16 CsI-coated photocathodes were replaced
by detectors consisting of arrays of MAPMTs coupled to individual fused silica lens telescopes.

In parallel, an extensive R\&D program~\cite{THGEM_rd}, aimed to develop MPGD-based large area PDs, established
a novel hybrid technology combining MicroMegas (MM)  and THick Gas Electron Multipliers (THGEMs)

In 2016 COMPASS RICH-1 was upgraded by replacing other 4  MWPCs-based PDs with novel detectors,
resulting from the newly developed MPGD hybrid technology~\cite{upgradeHybrid}. The description of
all aspects of this upgrade is the purpose of the present article.

\section{The development of THGEM-based PDs.}

MPGD-based PDs had previously been developed and successfully used by PHENIX-HBD~\cite{Anderson-2011-HBD},
in the form of triple GEMs operating in windowless mode, in pure CF$_4$.
The wide wavelength range (108-200 nm) Cherenkov photons are converted by the CsI layer on the first GEM
and large (6.2 cm$^2$) readout pads provide collective signals from several (5 to 10) photons.
To achieve an efficient detection of single photons in a narrower wavelength range a different approach,
based on the THGEM technology was chosen by the dedicated R\&D program for the COMPASS RICH upgrade
and also planned by the ALICE VMHPID project~\cite{VMHPID}.

THGEMs~\cite{THGEMs} are gaseous electron
multipliers derived from the GEM~design, scaling the geometrical parameters and
using standard Printed Circuit Boards (PCBs), where holes are obtained by drilling.
THGEMs with different geometrical parameters have been extensively characterized and
their performance as electron multipliers and as reflective photocathodes has been studied in
detail~\cite{THGEM_charact}.
THGEM-based PDs have been built and operated in Ar-based and in Ne-based gas mixtures:
gain values in the range of~10$^5$-10$^6$  and stable operation
have been obtained for small-size prototypes of various configurations~\cite{THGEM_gain}.
Time resolution values below 10~ns~\cite{THGEM_time} are typical.
Obtaining the same performance in terms of gain and stability with large or medium
size (300 $\times$ 300 cm$^2$) triple THGEM PDs was challenging:
a study of the origin of non-uniformity of the detector
response and the spark rates as well as the performance of different PD configurations
provided a specific procedure for large area THGEM production~\cite{THGEM_procedure},
quality assessment and configuration optimization. 
The Ion Back-Flow (IBF), which in a standard
triple THGEM configuration approaches 30\%, can be reduced by
a complete misalignment of the holes in different layers and using unbalanced values
of the electric field in the transfer regions between THGEMs~\cite{IBF}
and, more efficiently, by using a MM as last gas amplification stage: this hybrid
MM and THGEM detector architecture was tested and provided better
results in terms of PD performance and stability for large area
prototypes~\cite{THGEM-Hybrid-test} and was thus chosen for the COMPASS RICH-1 upgrade.

\section{The COMPASS hybrid PD architecture and components}

Each of the four new COMPASS PDs covers an active area of $\sim$ 600$\times$600 mm$^2$
and is formed by two identical modules ($600 \times 300$~$mm^2$), arranged side by side.

\begin{figure}[!htb]
\begin{minipage}[c]{.49\linewidth}
      \includegraphics[scale=1.01]{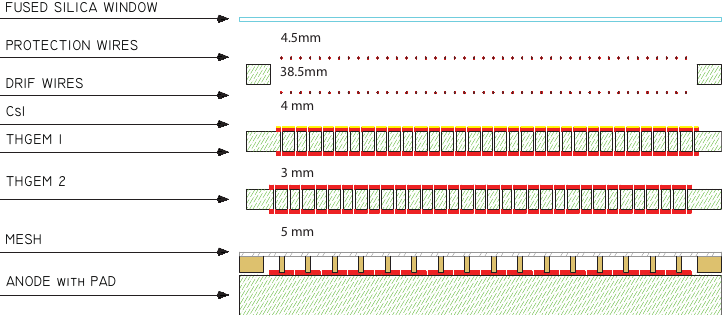}
\caption{Sketch of the hybrid single photon detector (image not to scale).}
\label{fig:hybridscheme}
   \end{minipage} \hfill
   \begin{minipage}[c]{.47\linewidth}
      \includegraphics[scale=0.24]{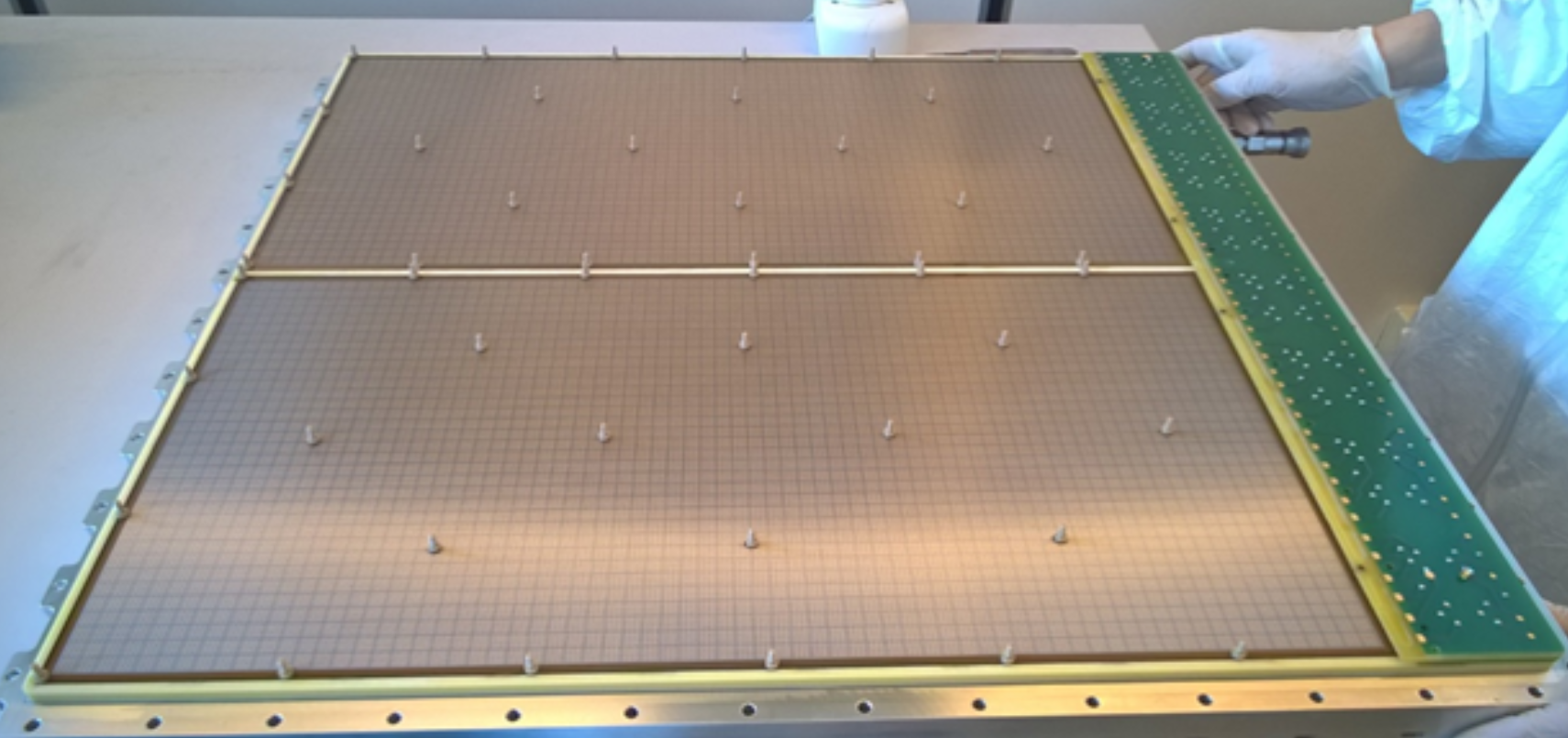}
\caption{Two Micromegas mounted side by side.}
\label{fig:Micromegas}
   \end{minipage}
\end{figure}

The hybrid module (Fig. \ref{fig:hybridscheme})
consists of two planes of wires, two layers of THGEMs and a
Micromegas on a pad-segmented anode.
UV light sensitivity is provided by a 300 nm thick CsI layer on the top
of the first THGEM electrode, which acts as a reflective photocathode for VUV photons.
A 600$\times$600 mm$^2$, 5 mm thick fused silica window separates the PD gas
volume from the RICH radiator volume.

The wire planes, called Protection and Drift, are made of $100~\mu$m diameter
Cu-Be, Au-coated wires, $\sim$600 mm long, with a pitch of 4~mm. 
The Protection wire plane is located at 4.5~mm from the fused silica window, it is
set at ground potential and collects the electrons from the volume above the THGEMs;
the Drift wire plane is placed at 4~mm from the CsI coated THGEM (and 38.5 mm from the
Protection plane) and is biased to the voltage which maximizes the extraction and collection of
the photoelectrons while repelling the ionization electrons.

All THGEMs (Fig. \ref{fig:THGEMs}, left) have the same geometrical parameters:
they are 470 $\mu$m thick (400 $\mu$m dielectric and
2 $\times$ 35 $\mu$m Cu),  581~mm long and 287~mm wide;
their holes have 400 $\mu$m diameter,
800 $\mu$m pitch and no rim. Holes located along the external borders have
500 $\mu$m diameter.
The top and bottom electrodes of each THGEM are segmented in 12 parallel sectors
(24 mm wide, apart from the border ones, which are 17 mm wide), separated by 0.7~mm
clearance area. 
Each sector of the THGEMs is electrically decoupled from the others by  a $1G\Omega$ resistor;
six consecutive sectors, grouped together, are fed by a specific high voltage power
supply channel. 
The two THGEM layers are mounted at a distance of 3 mm, in a configuration of
complete hole misalignment, to achieve the maximum charge spread;
a 5 mm gap
separates the second THGEM from the MM (Fig. \ref{fig:hybridscheme}).

The Micromegas (Fig. \ref{fig:Micromegas}) were produced at CERN using the bulk
technology procedure: they have a 128 $\mu$m gap, 18 $\mu$m woven stainless steel wire
mesh with 63 $\mu$m pitch, tensioned at about 10 N. A square array of 300 $\mu$m diameter pillars,
with 2 mm pitch, keeps the micromesh in place, on a PCB
specifically designed in Trieste for the COMPASS RICH-1 upgrade.

The anode PCBs are 3.2~mm thick, organized in 5 layers; they were produced by TVR srl
(Vicenza, Italy). After the MM production, they are cut to be 586 mm long and 283 mm wide.
The square anode pads facing the micromesh have 8 mm pitch and 0.5 mm inter-pad
clearance and are biased at a positive voltage, supplied via individual 470 M$\Omega$
resistors (one for each of the 4760 anode pads); they are divided in two groups (of 2380 pads),
powered by two independent high voltage channels.
The micromesh, which is the only non-segmented electrode, is stably kept at ground potential,
having a significant portion of its surface embedded in conductive glue.
Inside the anode PCB, 4760 buried readout pads, facing the anode pads and separated by a
70~$\mu m$ thick FR4 layer from them, transmit the signals with an attenuation
of only $\approx 10 \%$, thanks to the large individual capacitive coupling ( 40pF)
to their corresponding anode pad. This configuration prevents damages to the front-end electronics
in case a discharge occurs in the MM, grants a minimal gain drop ($\approx 4 \%$) in the pads
neighboring the discharging one, no discharge propagation and a restoring time of the
nominal voltage of $\sim 20 \mu$s.
The MMs are glued in pairs onto the detector holder frame, side by side
(Fig. \ref{fig:Micromegas}) and equipped with readout connectors and bias resistors
for the anode pads.
The MM gain uniformity distribution, measured using a prototype detector and a $^{55}Fe$
source, in an $Ar/CO2$~$70/30$ gas mixture, showed a standard deviation of $\sim 5\%$.

\section{THGEMs production and quality assessment.}

The THGEMs were produced from halogen-free EM 370-5 (Elite Material Co, Ltd.) raw PCB foils.
The thickness of the foils (resized to a square of 800~mm $\times$  800~mm to cut out the less uniform borders)
was mapped (in $\sim$ 1300 points/foil) using a Mitutoyo EURO CA776 coordinate measuring machine:
A typical thickness distribution for a good foil presents an average value of 472~$\mu$m and a 
standard deviation of 2 $\mu$m. 
Foils were accepted when $(th_{max}-th_{min}) \leq 15 \mu$m, where $th_{max}$ and $th_{min}$ 
are the maximum and minimum of the measured thickness values.

The PCB etching and the drilling of the holes ($\sim$300000 per THGEM) using a multi-spindle
machine (Posalux 6000-LZ) were performed by ELTOS S.p.A. (Arezzo, Italy).
The quality of the hole edge and walls was granted by replacing the drilling tool
every 1000 holes.
A specific post production procedure has been applied to all THGEMs~\cite{Polishing}
to round the hole borders and smooth the surface defects: it consisted in a careful electrode surface
polishing, using fine grain pumice powder, high pressure water rinsing and a mild chemical
attack by a SONICA PCB solution at 60~$^{\circ}C$ in ultrasonic bath, followed by a
thermal treatment.

\begin{figure}[!htb]
\includegraphics[scale=0.291]{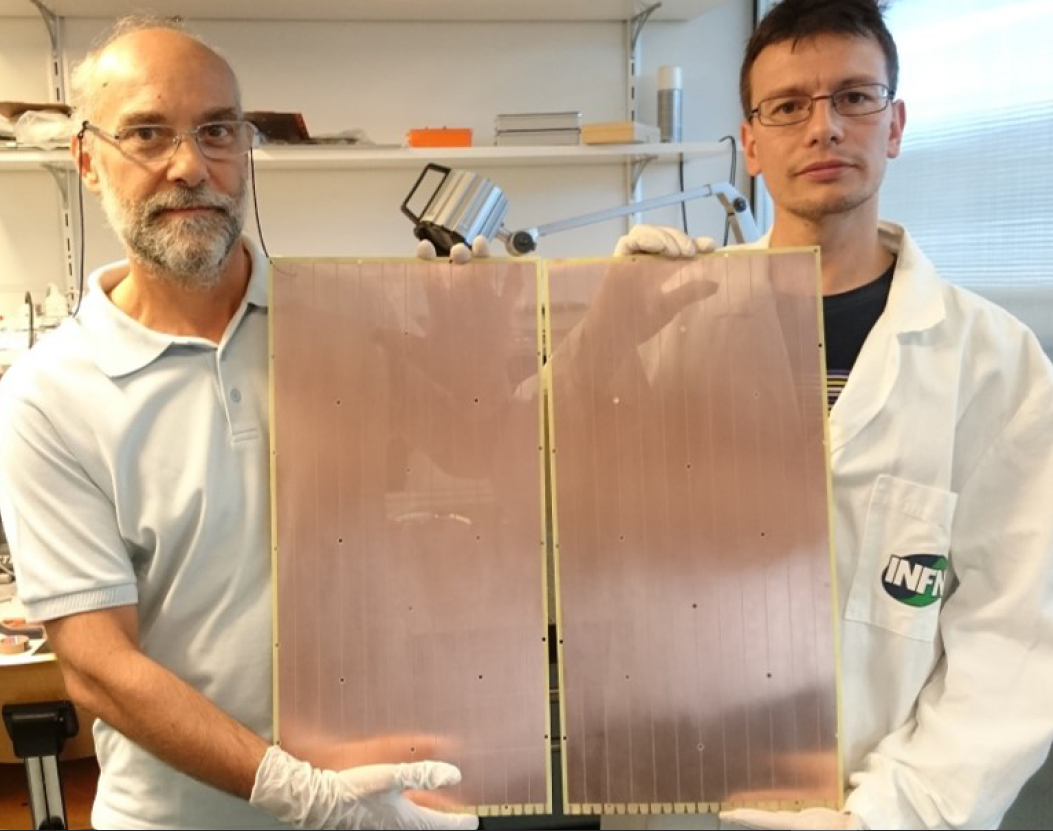}
\includegraphics[scale=0.262]{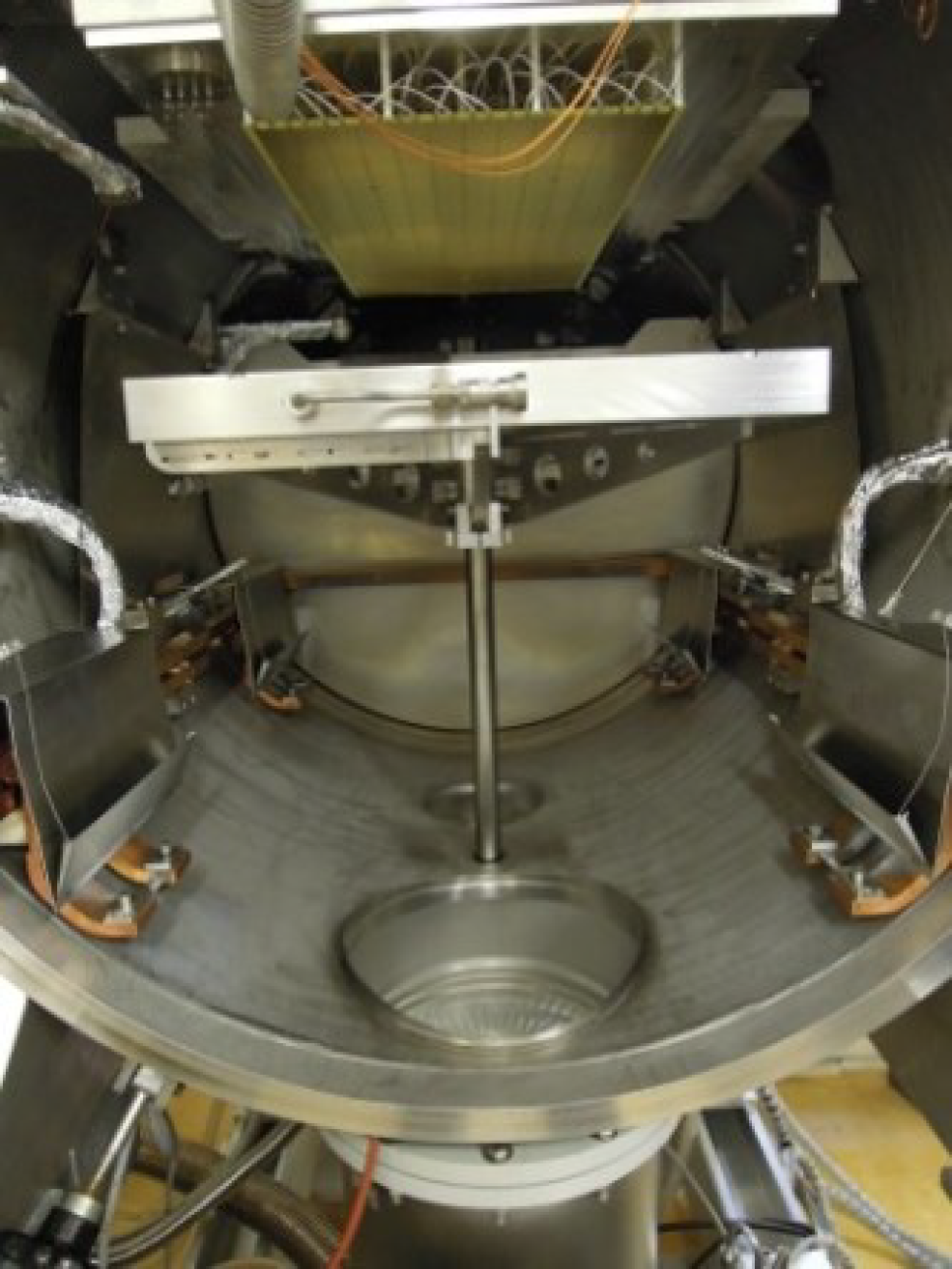}
\includegraphics[scale=0.225]{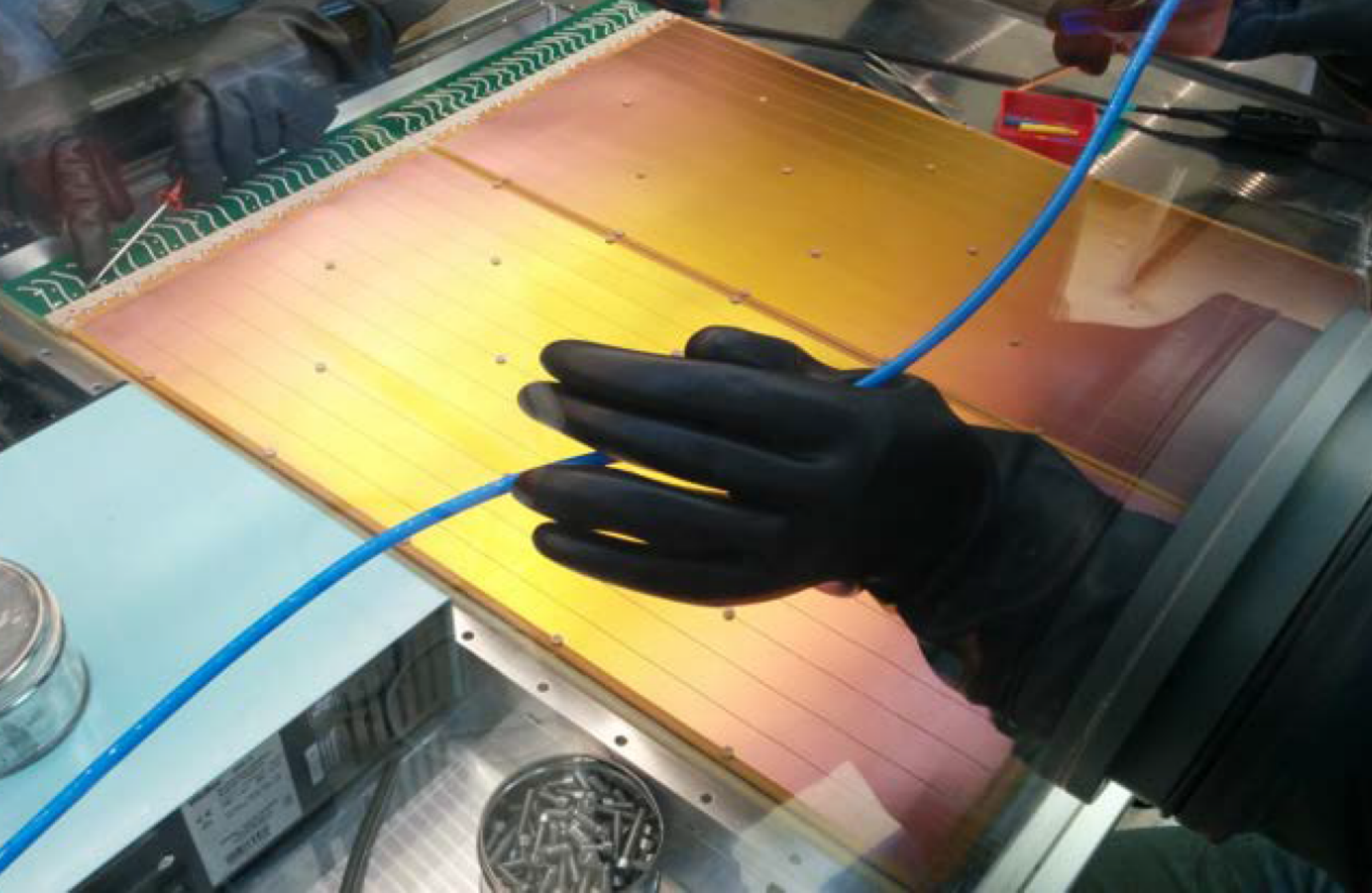}
\caption{Two THGEMs before the post-production treatment (left), a THGEM inside the CsI
evaporation plant (center), THGEMs being mounted on a COMMPASS Hybrid PD inside a glove-box (right).}
\label{fig:THGEMs}
\end{figure}

The electric strength and the response of the THGEM under high bias voltage condition
were verified in a test setup flushed with an $Ar/CO_{2}$~$70/30$ gas mixture where
the spark rate was measured: the bias voltage of each sector was automatically increased in steps of 10~V
and the number of discharges (defined as events with current values larger than 50~nA)
was counted. A bias voltage value corresponding to a discharge rate $\le$ 1 event per hour
was defined as stable voltage and a THGEM was qualified as electrically stable when all its
segments provided a maximum stable voltage exceeding 1200~V (which corresponds to an
effective gain of $\sim$60). The post production surface treatment was repeated
on THGEMS which failed the first electrical stability validation test.
Electrically stable THGEMs were then checked for their gain uniformity imposing a validation
threshold ($\sigma$ $\le$ 10$\%$), which was passed by almost all pieces, thanks to the
strict thickness uniformity selection criteria previously applied to the raw PCB material.

Qualified THGEMs were then coated with a Ni-Au layer ($\sim$ 5~$\mu m$ thickness 
Ni and 0.5~$\mu m$ thickness Au, both chemically deposited at CERN), to preserve the Cu surface
from oxidation and to prepare it as an optimal substrate for the CsI photocathode. 

The 300 nm thick CsI layer was deposited at the CERN Thin Film  Laboratory,
(Fig. \ref{fig:THGEMs}, center) following the RD26 procedure~\cite{Braem-CsI}
(thermal evaporation of CsI at $\sim$1 nm/s, in a vacuum of 10$^{-6}$ mbar + 8 h at 60 $^\circ$C).
A systematic measurement of photocurrent was performed on 60 points for each coated THGEM
using VUV light from a D$_2$ lamp to map the response uniformity after the CsI evaporation: 
good uniformity and small variations between different pieces ($\sigma \sim 10 \%$) were
observed; the overall Q.E. values are compatible with those obtained from the reference sample
used for COMPASS and ALICE photocathode production~\cite{Braem-CsI}, taking into account the
THGEM optical transparency of 23$\%$.

\section{Installation and commissioning}

The Hybrid PDs were assembled, equipped and tested before installation;
the correct position and planarity of the THGEMs inside the detector is guaranteed by 12 PEEK
pillars glued onto corresponding bases of photosensitive material prepared for them on the MM.
Auxiliary lateral electrodes, embedded in the chamber frames,
are used to correctly shape the electric field at the borders of the chamber volume.

The THGEM photo-cathodes were mounted using a large glove-box built for this purpose.
The four MWPC-based PDs to be replaced were part of a unique detector hosting 144 MAPMTs
and fused silica lens telescopes too: the MWPCs with CSI photocathodes were dismounted
from the RICH-1 vessel and the MAPMT systems were transferred onto the frames of the new hybrid
detectors. The combined PDs
were installed on COMPASS RICH-1 and equipped with front-end electronics, low voltages,
high voltages and cooling services during Spring 2016. 

The signals from the readout pads are collected by front-end electronic cards~\cite{APV} designed for
COMPASS RICH-1: they host four APV25-S1 chips, each reading 108 pads. Three front-end cards
are connected to a 10-bit flash ADC digitizer board equipped with a FPGA performing on-line zero
suppression. Data are registered by the COMPASS DAQ~\cite{COMPASSDAQ} and stored for offline analysis.
A cooling system using under-pressure water flow assures effcient removal of the heat
produced by the readout.

The hybrid PDs operate with an Ar/CH$_4$ 50/50 gas mixture; the
field configuration used in 2017 is: 0.4 kV/cm in the Drift region, 1.0 kV/cm in the two
transfer regions between the first and second THGEM and between the
second THGEM and the MM. The typical bias applied to the active
elements are: $\sim$ 620 V to the MM anode, $\sim$1200 V between top and bottom
of THGEMs.

\begin{figure}
	\centering
	\includegraphics[scale=0.236]{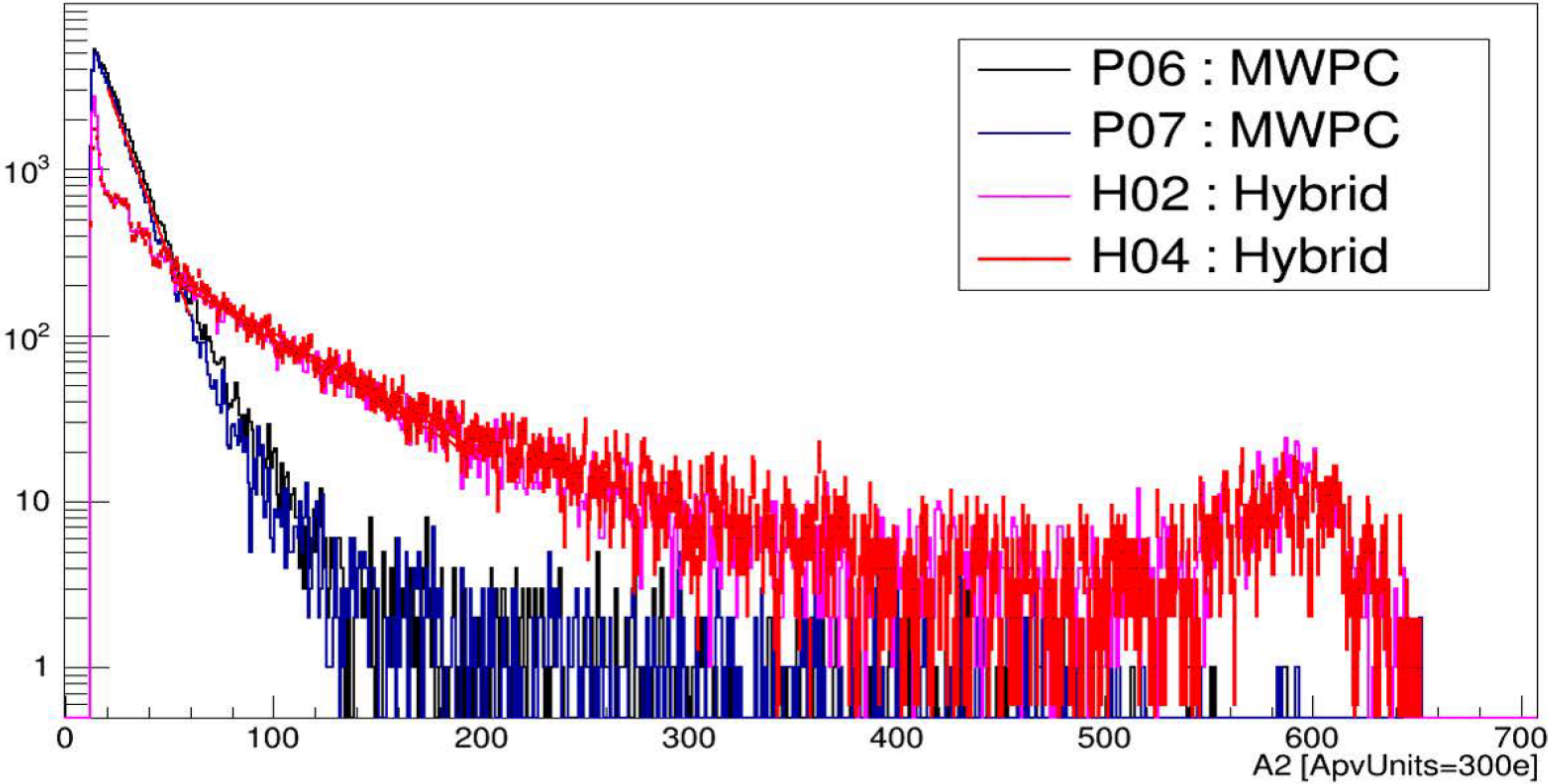}
	\includegraphics[width=0.43\linewidth]{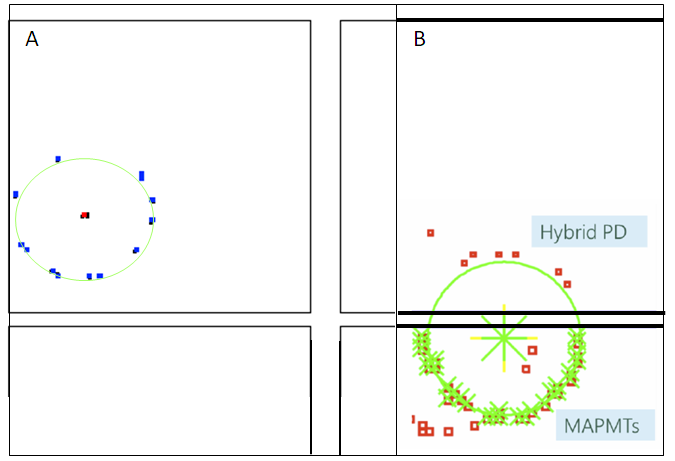}
	\caption{Signal amplitude distributions for MWPCs and Hybrids (left). Two Cherenkov rings from a typcal RICH-1 event in COMPASS 2017 data taking (right).}
	\label{fig:rings}
\end{figure}

The typical effective gain value is about 2$\times$10$^4$, larger than the gain provided
by the MWPC-based PDs, as can be seen in Fig. \ref{fig:rings}, left.

A high voltage monitoring program~\cite{Hybrid-HV}
stabilizes the gain by tuning the biases applied
to MM and THGEMs to compensate for the environmental changes of temperature and pressure:
gain variations, which would be as large as a factor of 2, are limited to $\sim$ 5$\%$ over
long running periods thanks to this compensation.
The IBF to the photocathode has been measured to be $\sim$ 3\%.

\par
The COMPASS hybrid PDs have been commissioned in 2016 and provide stable
performance in the 2017 run: an example of their
Cherenkov rings is presented in Fig. \ref{fig:rings}, right.
The configuration of the capacitive-resistive anode pads allow to operate the MM without
inconveniences even in case some anode pads are in short toward ground:
this condition was found for two pads during the Hybrid PD tests before installation and
for more pads during the commissioning run in 2016 and at the beginning of the 2017 run;
this however results in a dead area < 0.1$\%$ of the active area.

Before the 2017 run a refurbishing of the grounding distribution system for the Hybrid PDs
was performed: now the typical electronic noise for the new detector channels is $\sim$ 800
equivalent electrons r.m.s., stable at a few $\%$ level.

A preliminary evaluation of the number Cherenkov photons detected by the new Hybrid PD's
shows an increase with respect to the previously used MWPC-based PDs;
the full characterization of the new detectors and the upgraded COMPASS RICH-1 is still ongoing,
but indications of a stable and efficient performance of the hybrid PDs are clearly appearing.

The validity of the MPGD-based PD approach for RICH applications is fully confirmed by
the successful operation of COMPASS hybrid PDs.

 \acknowledgments
The activity is partially supported by the H2020 project AIDA2020 GA no. 654168.

\end{document}